# Optical Constants of Titan Haze Analogue from 0.4 to 3.5 µm: Determined Using Vacuum Spectroscopy

Chao He[1*], Sarah M. Hörst[1,2], Michael Radke[1], Marcella Yant[3],

[1] Department of Earth and Planetary Sciences, Johns Hopkins University, Baltimore, MD, USA che13@jhu.edu

[2] Space Telescope Science Institute, Baltimore, MD, USA

[3] Lockheed Martin Space, Littleton, CO, USA

Abstract

Titan's thick atmosphere is primarily composed of nitrogen and methane. Complex chemistry happening in Titan's atmosphere produces optically thick organic hazes. These hazes play significant roles in Titan's atmosphere and on its surface, and their optical properties are crucial for understanding many processes happening on Titan. Due to the lack of such information, the optical constants of laboratory prepared Titan haze analogues are essential inputs for atmospheric modeling and data analysis of remote sensing observations of Titan. Here, we perform laboratory simulations in a Titan relevant environment, analyze the resulting Titan haze analogues using vacuum Fourier transform infrared spectroscopy, and calculate the optical constants from the measured transmittance and reflectance spectra. We provide a reliable set of optical constants of Titan haze analogue in the wavelength range from 0.4 to 3.5 µm and will extend to 28.5 µm in the near future, which can both be used for analyzing existing and future observational data of Titan. This study establishes a feasible method to determine optical constants of haze analogues of (exo)planetary bodies.

# 1. INTRODUCTION

In Titan's atmosphere, complex chemistry induced by solar UV radiation and Saturn's magnetospheric electron bombardment produces thousands of organic molecules, leading to the formation of organic hazes and condensate layers (see e.g., Sagan et al. 1992; Coll et al. 1999, 2013; Cable et al. 2012; Hörst 2017). The organic materials will eventually fall to the surface of Titan and may interact with surface materials. The organic chemistry and resulting organic molecules are intriguing because of their prebiotic implications (see e.g., Hörst 2012; Neish et al. 2010; Sebree et al. 2018). Similar to Earth, Titan has a dense $N_2$-rich atmosphere, an Earth-like hydrological cycle, seasonal weather patterns, stable liquid on its surface (rivers, lakes, and seas), and other Earth-like surface features (sand dunes, deltas, and mountains), and most importantly, the complex, carbon-rich chemistry happening in the atmosphere and on the surface (Hörst 2017). The combination of all these features makes Titan an ideal place for studying prebiotic chemistry outside of Earth's environment. The successful Cassini-Huygens mission revealed many mysteries of Titan and rewrote our understanding of this Earth-like world, and the upcoming Dragonfly mission will explore the habitability and investigate prebiotic chemistry happening on Titan. The organic hazes not only affect Titan's physical and chemical processes, but also impact our observations and understanding of Titan. The optical and physical properties of Titan hazes are essential for modeling and analyzing observational data, but are not always available. Therefore, Titan haze analogues, or tholins, have been prepared in the laboratory to simulate chemical processes happening in Titan's atmosphere. Four decades of studies on tholins have advanced our understanding of Titan's atmosphere (see e.g., Coll et al. 2013; Cable et al. 2012; Brassé et al. 2015). The properties of tholins have been widely used for interpreting Titan observations, especially the optical constants that have been used for analyzing data acquired by several instruments onboard the Cassini-Huygens spacecraft (Tomasko et al. 2008; Bellucci et al. 2009; Rannou et al. 2010; Lavvas et al. 2010; Vinatier et al. 2012), including the Huygens Descent Imager/Spectral Radiometer (DISR: 0.35-1.7 µm), the Cassini Visible and Infrared Mapping Spectrometer (VIMS: 0.35-5.1 µm), and the Composite Infra-Red Spectrometer (CIRS: 7-1000 µm). The optical constants of tholins can also be utilized for future observations with the James Webb Space Telescope and the Dragonfly mission.

Several studies have reported the optical constants of tholins that were prepared using different setups and under different conditions (e.g., Khare et al. 1984; Ramirez et al. 2002; Tran et al. 2003; Vuitton et al. 2009; Hasenkopf et al. 2010; Sciamma-O'Brien et al. 2012; Imanaka et al. 2012; Mahjoub et al. 2012; Ugelow et al. 2017). They also used different techniques to determine the optical constants, such as spectrophotometry, ellipsometry, Brewster angle spectroscopy, Cavity Ring Down Spectroscopy (CRDS), and photo deflection spectroscopy. Unsurprisingly, Titan tholins from these studies show wide variations in optical constants (Brassé et al. 2015). The tholins used in the previous work were prepared at room temperature (300 K) rather than cryogenic temperature to simulate Titan's environment. Most of the optical measurements were performed under ambient conditions, and the results may be impacted by contaminations from Earth's atmosphere. The purpose of this study is to provide a reliable set of optical constants of the Titan tholins produced using the Planetary HAZE Research (PHAZER) chamber (He et al. 2017). Our Titan experiment is conducted at ~100 K and the optical properties are measured using vacuum Fourier transform infrared spectroscopy (FTIR). We focus on the optical constants in the wavelength range from 0.4 to 3.5 µm in this study and will extend to 28.5 µm in the near future.

## 2. MATERIALS AND EXPERIMENTAL METHODS

*2.1. Haze Analogue Production*

The Titan haze analogue is produced using the PHAZER chamber (Figure 1) at Johns Hopkins University (He et al. 2017). The gas mixture is prepared in a stainless-steel cylinder by mixing 5% $CH_4$ (99.99% Airgas) in $N_2$ (99.999% Airgas). The gas mixture is cooled down to 100 K by flowing through a cooling coil. As reported previously (He et al. 2017), the 15 m stainless steel cooling coil is immersed in liquid nitrogen (77 K), which cools the gas mixture and removes trace impurities in the gas mixture, and the temperature of the gas mixture is determined based on the ideal gas law ($P_1/P_2=T_1/T_2$). The gas flow rate is 10 standard cubic centimeters per minute (sccm) and the pressure in the chamber is 2 Torr. Under such conditions, the reactant gas mixture is exposed to the cold plasma discharge for about 3 seconds. The detailed experimental procedure can be found in our previous study (He et al. 2017).

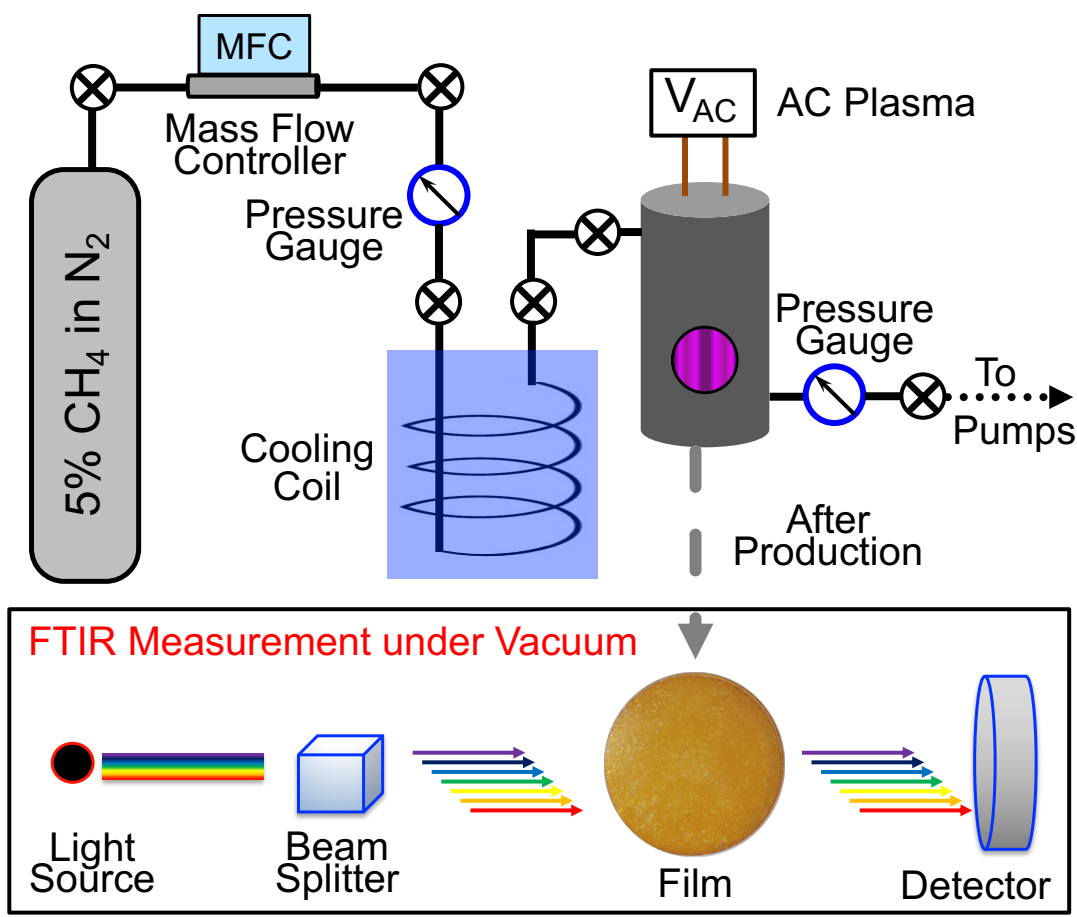

Figure 1. Schematic of the experimental setup for tholin film sample production and FTIR measurement.

To prepare a film sample for the spectroscopy measurement, we placed optical-grade quartz substrates (diameter: 25.4 mm, thickness: 1.6 mm, Ted Pella, Inc.) in the chamber for sample collection. After continually running the experiment for 72 hours, we put the chamber under vacuum for 48 hours to remove the volatile components. Therefore, our measurement may not be able to capture some spectral features of the volatile components in the tholin. After warming up, the film samples were collected in a dry-$N_2$ glove box (InertCorp.). The fresh produced film samples (within 2 hours after collection) are used for the spectroscopy measurement to avoid aging or contamination. The film samples are yellow orange in color as shown in Figure 1 and the surfaces are very smooth with surface roughness ($R_q$) less than 3 nm (examined using atomic force microscopy, He et al. 2018a).

*2.2 Vacuum Fourier-transform Infrared Spectroscopy (FTIR) Measurement*

The film samples deposited on quartz substrates are characterized with a Vertex 70v FTIR spectrometer (Bruker Corp.). It is a vacuum spectrometer (both the sample compartment and the optics), which can reduce the spectral features of Earth's atmosphere ($H_2O$ or $CO_2$ absorptions), and thus increase the peak sensitivity without masking very weak spectral features. The spectrometer is capable of measuring transmittance and reflectance from 0.4 to 28.5 μm (25,000 cm to 350 cm$^{-1}$) with a resolution of 0.4 cm$^{-1}$. Our film samples are deposited on quartz substrates because quartz is chemically inert, neither changing under glow discharge nor reacting with the gas mixture in our reaction chamber. However, quartz is opaque to wavelengths longer than 3.5 μm. Therefore, in the current study, we only measured the film samples on quartz substrates in the wavelength range of 0.4 to 3.5 μm despite the spectrometer's wider wavelength coverage. We will explore other potential substrates that can be used in the reaction chamber and also transparent in mid-IR range, such as $CaF_2$, KBr, NaCl, and Si, in order to accurately measure the spectra of the film samples in the mid-IR range.

For the measurement, a fresh film sample was carefully transferred to the spectrometer for measurement without exposure to air (using a $N_2$ glove bag). The transmittance and reflectance of the film on quartz substrate were measured from 0.4 to 3.5 μm under vacuum (below 0.2 mbar) at room temperature (294 K). The spectra were acquired with two detectors (silicon diode and DLaTGS detector) and two beamsplitters (quartz beamsplitter and KBr beamsplitter). From 0.4 to 1.11 µm (25000 to 9000 cm$^{-1}$), we employed the silicon diode detector and quartz beamsplitter; from 0.83 to 1.25 µm (12000 to 8000 cm$^{-1}$), we employed the DLaTGS detector and quartz beamsplitter; and from 1.11 to 3.5 µm (9000 to 2850 cm$^{-1}$), we employed the DLaTGS detector and KBr beamsplitter. Overlapping data confirm that the spectrometer is calibrated properly across different wavelength ranges. We configured the instrument to acquire 250 scans for each measurement and to obtain spectra with a resolution of ~2 cm$^{-1}$. The transmittance was measured at normal angle with standard sample holder, while the reflectance was measured with a reflection accessory (A510 reflection unit). We measured the transmittance of a blank quartz disc and the reflectance of the gold standard as references. The transmittance and reflectance spectra of the sample were obtained by dividing the sample measurement by the reference measurement in each case. This procedure ensures that all other factors (the light source, the beamsplitter, the mirrors, and the detector) affecting the spectrum are eliminated and the resulting spectrum only displays the spectral features originating from the sample itself. Note that the true reflectance of our sample from 0.4 to 0.7 µm can be up to 64% lower, because we use a gold mirror as reference, which is not calibrated, and the reflectance of the gold mirror drops significantly below 0.7 µm (0.97 at 0.7 µm, 0.93 at 0.65 µm, 0.9 at 0.6 µm, 0.78 at 0.55 µm, 0.45 at 0.5 µm, and 0.36 at 0.4 µm) (Bennett & Ashley 1965; Beran 1985). We discuss how the uncertainty of the reflectance spectrum affects the derived optical constants in Section 3.1.

*2.3. Spectroscopy Data Analysis*

To calculate n and k from the transmittance (T) and reflectance (R) of the film sample, we need to know the film thickness. For a thin film, light waves reflected from upper and lower surfaces will interfere with one another, either enhancing or reducing the reflected light. The film thickness (*t*) can be calculated from the appearance of the interference fringe in the spectrum by Eq. 1 (Rancourt 1996; Stenzel 2005).

$$t = \frac{1}{2\sqrt{n^2 - sin^2 i}} \times \frac{x}{(v_1 - v_2)} \qquad (Eq.\ 1)$$

where *n* is the refractive index of the film material; *i* is the angle of incidence; *x* is the number of fringes between two wavenumbers ($v_1$ and $v_2$).

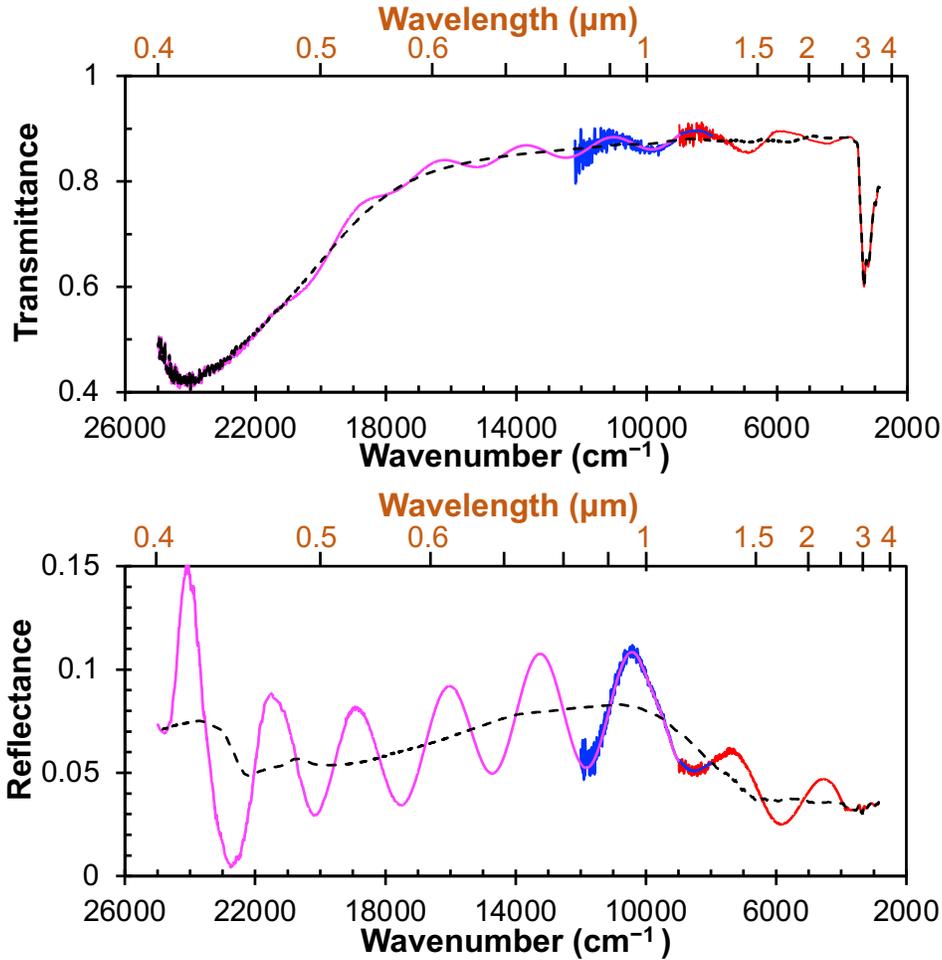

Figure 2. The transmittance (top) and reflectance (bottom) spectra of our Titan tholin film sample in the visible and near IR region (25000 to 2850 cm$^{-1}$, or 0.4 to 3.5 μm). The spectra in red (2850 to 9000 cm$^{-1}$) were acquired with the DLaTGS detector and KBr beamsplitter, the spectra in blue (8000 to 12000 cm$^{-1}$) with the DLaTGS detector and quartz beamsplitter, and the spectra in purple (9000 to 25000 cm$^{-1}$) with the silicon diode detector and quartz beamsplitter. The interference fringes are observed in the original spectra and the spectra after removing the interference fringes are shown in the black dashed lines.

Figure 2 shows the transmittance and reflectance spectra of our film sample. Interference fringes are observed in both spectra. From the observed interference fringes in the reflectance spectrum, we can get $x$, $v_1$ and $v_2$ for Eq. 1; the angle of incidence ($i$) is 11° in our experiments. The refractive index of our tholin sample is unknown at this point. However, several studies have reported that the refractive index of tholin samples are about 1.5–1.7 in the wavelength range from 0.4 to 1.2 μm. So, we first estimated the film thickness by using an average reported value ($n_o$=1.55) of the refractive index. The estimated film thickness ($t_1$) is 1.5 μm. We re-calculated the film thickness after we constrained the refractive index of our tholin sample. As the choice of initial n value, we tried different initial n values ($n_o$=1.50, 1.55, 1.60, or 1.70) for the calculation and found that the initial choice of the n value only increases or decreases the times of the calculation cycle but does not affect the final result.

The interference fringes were removed following the method described in Neri et al. (1987) to eliminate their effect on optical constant calculation. The transmittance and reflectance spectra after the interference fringe removal are shown in Figure 2 (dashed lines). The correction was given in Neri et al. (1987) as

$$F(x_n) = \frac{2G(x_n)+G(x_{n+m})+G(x_{n-m})}{4} \qquad \text{(Eq. 2)}$$

where $x_n$ is the nth abscissa, $F(x_n)$ is the fringe-removed spectrum value at $x_n$, $G(x_n)$ is the original spectrum value at $x_n$, $G(x_{n+m})$ and $G(x_{n-m})$ are the original spectrum values at shifted abscissae and $2m$ is the maximum integer number of points contained in the interval $d$, which is the average fringe spacing. For our transmittance and reflectance spectra, the average fringe spacing ($d$) is about 2800 cm$^{-1}$ and there are 1448 points ($m$=724) contained in the interval $d$.

The reflectance and transmittance spectra after the interference fringe removal are plotted separately in Figure 3 as a function of wavelength, which shows the spectral features more obviously. As shown on the transmittance spectrum, there are a few absorptions features at 0.42, 3.0, 3.1, 3.36, 3.40, and 3.48 μm. The absorption at 0.42 μm is consistent with the yellow color of the film, likely caused by the presence of aromatic compounds and/or unsaturated species with conjugated pi bonds (Rao 1975; van Krevelen & te Nijenhuis 2009). The peaks at 3.0, 3.1, 3.36, 3.40, and 3.48 µm, are caused by absorptions of organic functional groups in the sample. The absorptions at 3.0 (~3330 cm$^{-1}$) and 3.1 µm (~3230 cm$^{-1}$) are characteristic of stretching of N-H bonds, while the features at 3.36 to 3.48 µm (2870 to 2980 cm$^{-1}$) are due to C-H bond stretching (Lin-Vien et al. 1991; Socrates et al. 2001). The absorption features of N-H and C-H bonds indicate the presence of amines and hydrocarbons in the sample, which is consistent with our previous result with nuclear magnetic resonance spectroscopy (NMR) on the sample produced under identical conditions (He et al. 2017). Other functional groups, such as nitriles, imines, and aromatics (e.g., Coll et al. 1999; Imanaka et al. 2004; He et al. 2012; Gautier et al. 2012; Sciamma-O'Brien et al. 2017), are likely present in the sample, but their absorption features are beyond the wavelengths we measured.

We calculated the optical constants ($N=n+ik$, where $N$ is the complex refractive index, $i$ is the square root of -1, $n$ is the refractive index, and $k$ is the extinction coefficient) of the film following the method described in Khare et al. (1984). Because the quartz substrate is transparent in the studied wavelength range (0.4 to 3.5 µm), the transmittance ($T$) can be expressed as in Eq. 3.

$$T = (1-R)e^{-4\pi kt/\lambda} \quad \text{(Eq. 3)}$$

where $R$ is the reflectance, $t$ is the film thickness, and $k$ is the extinction coefficient. Then, the extinction coefficient ($k_1$) can be calculated with the estimated thickness and the measured transmittance and reflectance spectra (Eq. 4).

$$k = \frac{\lambda}{4\pi t} \ln\frac{(1-R)}{T} \quad \text{(Eq. 4)}$$

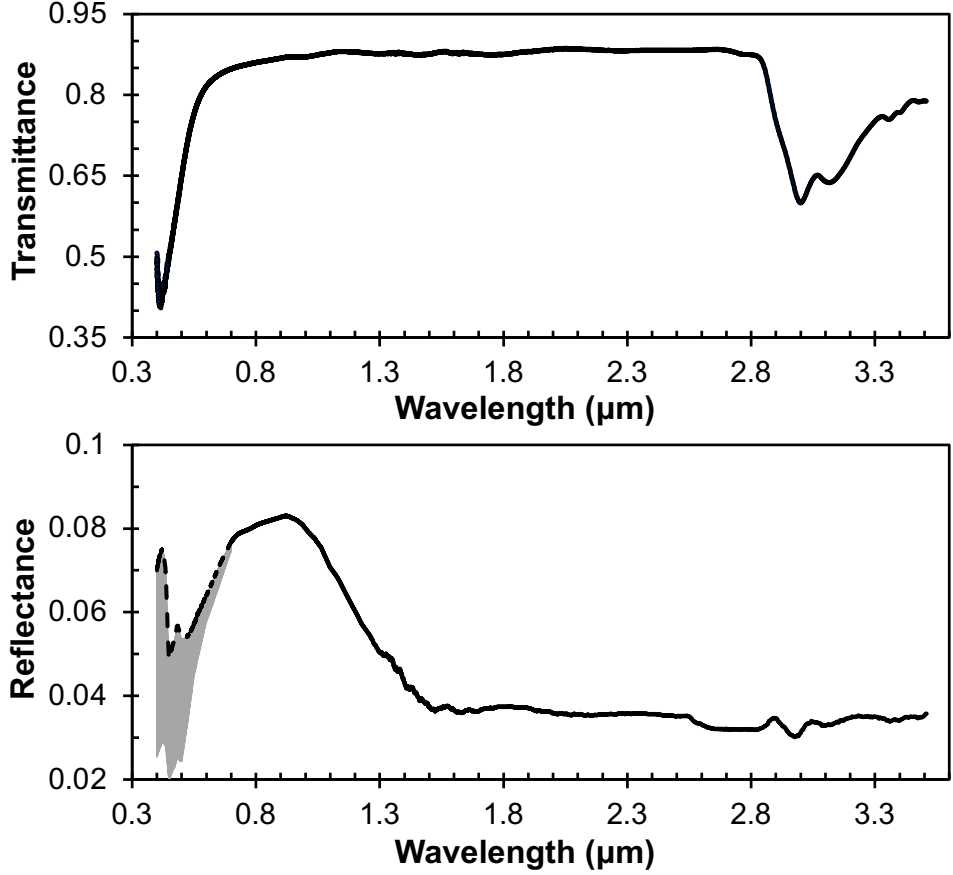

Figure 3. The transmittance (top) and reflectance (bottom) spectra of our Titan tholin film sample as a function of wavelength after removing the interference fringes. The reflectance from 0.4 to 0.7 μm is shown as dashed line with shaded gray area for the uncertainty due to the uncalibrated gold mirror).

According to the Fresnel equation for the reflection coefficient ($r$) from an absorbing medium, the reflectance ($R$) can be expressed as (Jahoda 1957; Khare et al. 1984; Mistrik et al. 2017)

$$R = \frac{(n-1)^2+k^2}{(n+1)^2+k^2} \quad \text{(Eq. 5)}$$

Then, the refractive index (n) can be determined from the reflectance data ($R$) and calculated extinction coefficient ($k$) using the following relation

$$n = \frac{1+R+\sqrt{4R-k^2(1-R)^2}}{1-R} \quad \text{(Eq. 6)}$$

Note that Eq. 5 is only an approximation to the reflectance measured in this study, because Eq. 5 does not take the properties of the substrate or the film thickness into account; this simplified equation allows an analytical solution for the value of n with Eq. 6. Using the new calculated refractive index ($n_1$) of our sample, we calculated a new set of the film thickness ($t_2$), the extinction coefficient ($k_2$), and the refractive index ($n_2$) with Eq. 1, Eq. 4, and Eq. 6. We continued the calculation cycle until the change of the refractive index is less than 0.2% [($n_{m-1}-n_m$)/$n_m$<0.002]. This procedure was done at all wavelengths. Based on Eq. 6, the n value is not sensitive to the change of k when k is very small. In our last calculation cycle, the change of the n values is very subtle (<0.01%) from 0.4 to 2.8 µm and is slightly larger (0.01-0.2%) from 2.8 to 3.5 µm. The largest change (0.2%) occurs at 3.0 µm. The final thickness was determined to be 1.219 µm, while the final extinction coefficient ($k_m$) and refractive index ($n_m$) is shown in Figure 4. Note that we used the reflectance measured at near-normal incidence angle (11 degrees) instead of the reflectance at normal incidence angle, because the values should be very close to each other. According to the Fresnel equations and Schlick's model (Schlick 1994), the specular reflection coefficient R can be approximated by:

$$R = R_0 + (1 - R_0)(1 - cos\theta)^5 \qquad (Eq.\ 7)$$

where $R_0$ is the reflection coefficient at normal incidence, and θ is the incidence angle. For our case, the incidence angle is 11 degrees, and $(1-cos\ \theta)^5$ is about $2*10^{-9}$. So, the last term in the equation can be neglected and the reflection coefficient at near-normal incidence (11 degrees) is very close to the reflection coefficient at normal incidence, i.e., $R=R_0$.

Ramirez et al. (2002) showed that light scattering could affect the calculation of optical constants. When light scattering is significant, the calculations performed with specular measurements would lead to an underestimation of the extinction coefficient (k). The scattering loss ($I_s$) on a surface is related to surface roughness ($R_q$) and light wavelength (λ), following Eq. 8 (Bennett & Porteus 1961), where $R_0$ is the reflectance of a perfectly smooth surface.

$$I_s = R_0[1 - e^{-(4\pi R_q/\lambda)^2}] \qquad (Eq.\ 8)$$

The scattering loss ($I_s$) decreases toward longer wavelengths. The surface of our tholins films is very smooth, with surface roughness ($R_q$) less than 3 nm as demonstrated with atomic force microscopy (He et al. 2018). Therefore, the scattering loss ($I_s$) on our tholin film is less than 0.8% in the wavelength range we measured, which has a negligible effect (within the measurement error) on the optical constant calculations with the specular measurements.

Besides using Eq. 6, we obtained another set of n values independently by performing the Kramers–Kronig analysis. The Kramers-Kronig relation between n and k is given by the dispersion relation (see, e.g., Wood and Roux 1982; Toon et al. 1994; Imanaka et al. 2012):

$$n(\nu) = n_0 + \frac{2(\nu^2 - \nu_0^2)}{\pi} P \int_0^\infty \frac{\nu' k(\nu')}{(\nu'^2 - \nu^2)(\nu'^2 - \nu_0^2)} d\nu' \quad \text{(Eq. 9)}$$

where ν represents wavenumber (cm$^{-1}$), P indicates the Cauchy principal value, and $n_0$ is the real refraction index at $\nu_0$. The principal value should be integrated for the entire wavelength range. We used our derived k values from 2850 to 25000 cm$^{-1}$ in the integrand and assumed that the k values outside the measured wavelength range is a constant. The assumption for the integral in the Kramers-Kronig relation (Eq. 9) is generally valid unless a large local absorption peak exists just outside our measurement range. As discussed in previous studies (see, e.g., Toon et al. 1994; Imanaka et al. 2012), using an anchor point ($n_0$) can reduce the uncertainty from the above numerical integration. Here, we used the n value determined from Eq. 6 at 2.91281 μm ($\nu_0$=3333.11 cm$^{-1}$) as the anchor point because its uncertainty is relatively small (~2.7%). By employing the same calculation cycle procedure, we calculated a new set of n, k, and the film thickness (t) with Eq. 1, 4, and 9. The iterative process stops when the termination criteria is met [$(n_{m-1}-n_m)/n_m<0.002$]. We also calculated using a different anchor point (n value at 1.36802 μm determined from Eq. 6) and found that the calculations using two different anchor points (1.36802 μm VS. 2.91281 μm) yield similar results (the differences are within the range of uncertainties). We present the results from the calculation using the anchor at 2.91281 μm because the uncertainties are smaller (Figure 4 and Supplementary Table 1). From this method, the thickness is determined to be 1.293 μm. Due to the small thickness difference, the derived k values are ~6% lower than those derived from Eq. 1, 4, and 6. However, the k values are not plotted separately in Figure 4 because the change is uniform and small. The derived n values from the Kramers–Kronig analysis (Eq. 9) are shown in Figure 4 for comparison with those derived from Eq. 6.

## 3. RESULTS AND DISCUSSIONS

*3.1. Optical constants of our Titan tholins compared with other Titan tholins*

Fig. 4 shows the optical constants of our Titan tholin sample in the wavelength from 0.4 μm to 3.5 μm, along with previously reported optical constants of Titan tholins in this wavelength range (Khare et al. 1984; Ramirez et al. 2002; Tran et al. 2003; Vuitton et al. 2009; Sciamma-O'Brien et al. 2012; Imanaka et al. 2012; Mahjoub et al. 2012; Ugelow et al. 2017). In the measured wavelength range, the extinction coefficient, *k*, varies greatly from 0.0037 to 0.093. The two sets of refractive indices (n) obtained from two independent methods (Eq. 6 and 9) are close. The n values are in the range between 1.41 and 1.76, and the value differences from two methods are up to 14% at short wavelength (0.4 to 1.5 μm) but less than 5% at longer wavelength (>1.5 μm).

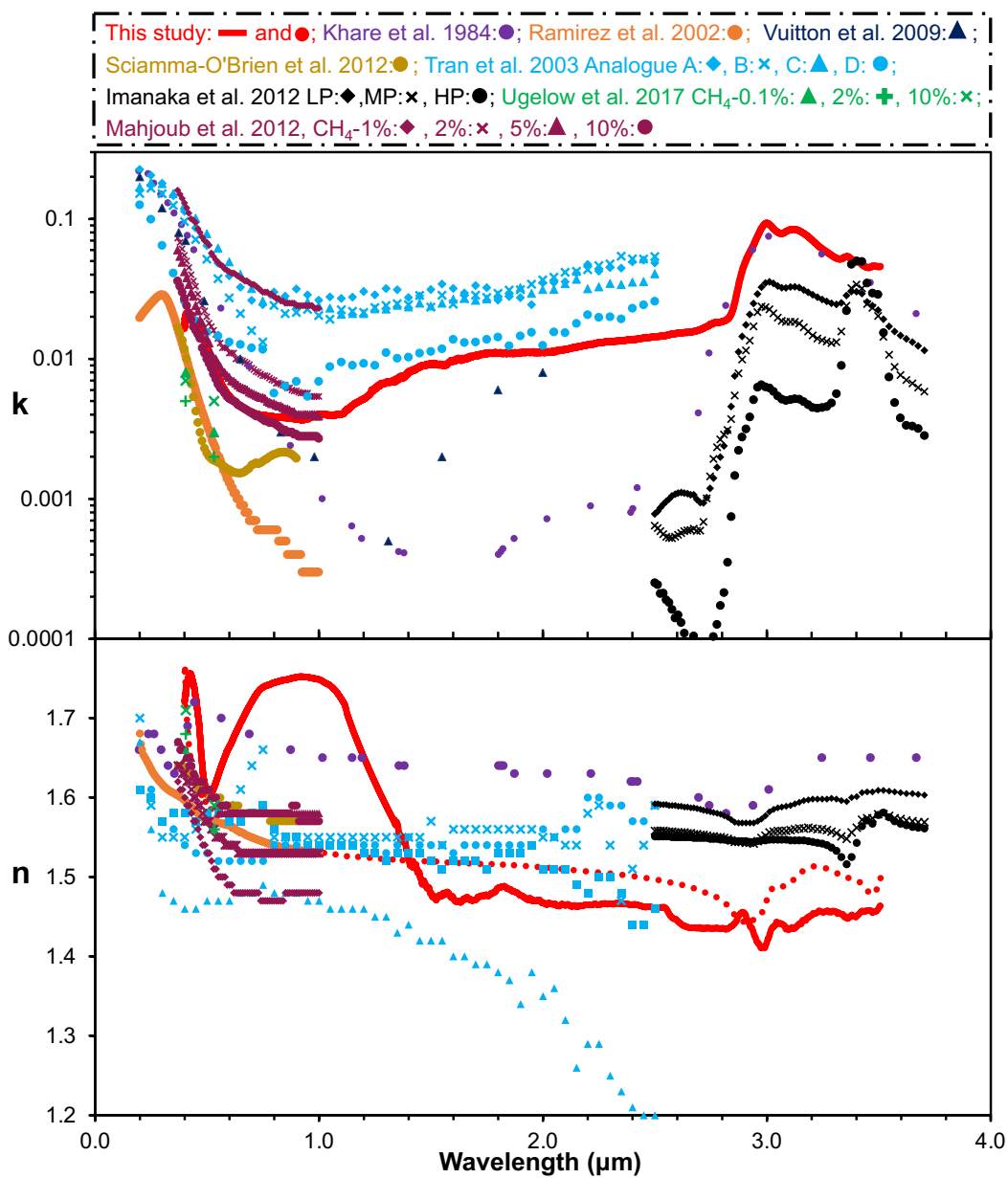

Figure 4. The optical constants [Top: extinction coefficient (k); Bottom: refractive index (n) obtained from two independent methods (solid red line from Eq. 6 and dotted red line from Eq. 9)] of our Titan tholin sample (red) from 0.4 μm to 3.5 μm. The optical constants of Titan tholins in this wavelength range from previous studies are also plotted for comparison (Color and symbols are labeled on the top). There are several factors which introduce uncertainties to the derived optical constants, including the systematic measurement uncertainty from the spectrometer, the large uncertainty of the measured reflectance below 0.7 μm due to the gold mirror, the interference fringe removal, and the film thickness calculation. In addition, the uncertainties of the derived k values propagate to the derived n values in both methods (Eq. 6 and 9), and the uncertainty of the anchor n value (from Eq. 6) also propagate to the n values derived from Eq. 9. We evaluated all the factors and estimated the overall uncertainties for the two sets of k and n (listed in supplementary Table 1 along with the n and k values). The two sets of k values obtained from the two methods have similar uncertainties. The uncertainties for the k values are about 3-6% at longer wavelengths (>1.5 µm) but increase to 10% at short wavelengths (0.4 to 1.5 µm). The two sets of n values also have small uncertainties (less than 4%) at longer wavelengths (>1.5 µm). At short wavelengths (0.4 to 1.5 µm), the n values derived from the first method (Eq. 1, 4, and 6) have larger uncertainties (5-14%), while the n values calculated from the Kramers–Kronig method (Eq. 1, 4, and 9) have smaller uncertainties (3-6%) for two reasons. First, the k values are integrated over the entire wavelength range in Eq. 9, so the larger uncertainties of the small k values in this range carry less weight when calculating n values with Eq. 9; second, Eq. 9 does not use the measured reflectance directly, so the large uncertainty of the measured reflectance below 0.7 µm has little impact on the calculated n value. Therefore, we believe the n values obtained from the Kramers–Kronig analysis (Eq. 9) are closer to the real n values at short wavelength (0.4 to 1.5 µm). The comparison with prior studies in Figure 4 confirms that the value and the trend of the n values obtained from the Kramers–Kronig analysis (Eq. 9) are more similar to the values reported previously. Nonetheless, the similarity between two sets of n values indicates that both methods provide relatively accurate n values, with the Kramers–Kronig analysis more accurate at short wavelength (0.4 to 1.5 µm).

The optical constants of Titan tholins in the range 2.5 to 3.5 µm are reported in two other studies (Khare et al. 1984 and Imanaka et al. 2012). In this range, the extinction coefficient (k) of our sample is comparable to that in Khare et al. (1984) from 2.8 to 3.3 µm, but higher at other wavelengths (2.5 to 2.7 µm and 3.3 to 3.5 µm). Compared to the data from Imanaka et al. (2012), our k values are always higher in the range of 2.5 to 3.5 µm. The N-H absorption features (3.0 and 3.1 µm) in our sample are comparable to those of Khare et al. (1984), but stronger than those of Imanaka et al. (2012); while the C-H features (3.36-3.48 µm) in our sample are comparable to that in the low-pressure sample of Imanaka et al. (2012), but stronger than all other samples in both studies. In the shorter wavelength (0.4 to 2.5 µm), our k values fall within the range of the reported values in previous studies (Khare et al. 1984; Ramirez et al. 2002; Tran et al. 2003; Vuitton et al. 2009; Sciamma-O'Brien et al. 2012; Mahjoub et al. 2012; Ugelow et al. 2017). As we discussed earlier, we think the n values obtained from the Kramers–Kronig analysis (Eq. 9) are closer to the real n values at short wavelength (0.4 to 1.5 µm). Our n values from both methods fall within the range of the reported values from 1.5 to 2.5 µm, but are at the lower end from 2.5 to 3.5 µm. We should note that the optical constants compared here were determined using different techniques, such as spectrophotometry (Khare et al. 1984; Ramirez et al. 2002; Tran et al. 2003; Imanaka et al. 2012), ellipsometry (Sciamma-O'Brien et al. 2012; Mahjoub et al. 2012), Brewster angle spectroscopy (Khare et al. 1984), CRDS (Hasenkopf et al. 2010; Ugelow et al. 2017), Photo Deflection Spectroscopy (Vuitton et al. 2009), etc. In those studies, the tholin samples were also produced under different conditions using different setups.

The different experimental conditions and the techniques used to determine the optical constants can be found in Table 1 and Table 2 in Brassé et al. (2015), except a recent study by Ugelow et al. (2017). They produced tholins with 0.1%, 2%, or 10% of $CH_4$ in $N_2$ using spark discharge at room temperature, and determined the optical constants at 0.405 and 0.532 µm with CRDS technique. The gas mixtures used in the previous studies (Khare et al. 1984; Ramirez et al. 2002; Tran et al. 2003; Vuitton et al. 2009; Sciamma-O'Brien et al. 2012; Mahjoub et al. 2012; Ugelow et al. 2017) are primarily $N_2$ (>90%) and $CH_4$, with one study adding trace species (Tran et al. 2003). Two kinds of energy sources, UV photons and plasma discharges are used to induce the chemical processes. The tholin samples produced with electrical discharges usually have higher nitrogen contents than those with UV photons because the electrical discharges are energetic enough to dissociate $N_2$ directly (e.g., Imanaka et al. 2004; He et al. 2017).

Among the tholin samples compared here, our tholin sample is the only one that was produced at low temperature (~100 K) simulating the cold environment in Titan's atmosphere (He et al. 2017). We have also been extremely careful with sample collection (in $N_2$ glovebox) and measurements (under vacuum) to avoid contamination and eliminate the spectral features from Earth's atmosphere. Measurements in previous studies were usually done under ambient conditions or in a purged system. Among the optical constants compared here, only Khare et al. (1984) and our study have continuous data in this wavelength range (0.4 to 3.5 μm). However, the optical constants in this range from Khare et al. (1984) were calculated based on two measurements (0.4 to 2.5 μm, and >2.5 μm) on different instruments with different samples, furthermore these measurements were done with very low resolution. Our measurements here used one film sample on a single vacuum FTIR spectrometer. We will extend the measurement to mid-IR range (up to 28.5 μm) with the same spectrometer in the near future.

*3.2. Implication for Titan and beyond*

The optical constants of Titan tholin samples have been used for atmospheric modeling and observations data analysis for Titan (e.g., Rages & Pollack 1980; Tomasko et al. 2008; Bellucci et al. 2009; Rannou et al. 2010; Lavvas et al. 2010; Vinatier et al. 2012), in particular the data from Khare et al. (1984) has been widely used because of its extensive spectral coverage, from 0.025 μm to 1000 μm. We need to acknowledge that Khare's work provided fundamental optical data for many theoretical models and helped us gain valuable information on Titan's haze and atmosphere. Indeed, the optical data from Khare et al. (1984) could approximately fit the observations in visible wavelengths. However, we should also realize that Khare's optical data may not fit well with Cassini VIMS and CIRS observations in certain wavelengths (Bellucci et al. 2009; Rannou et al. 2010; Vinatier et al. 2012).

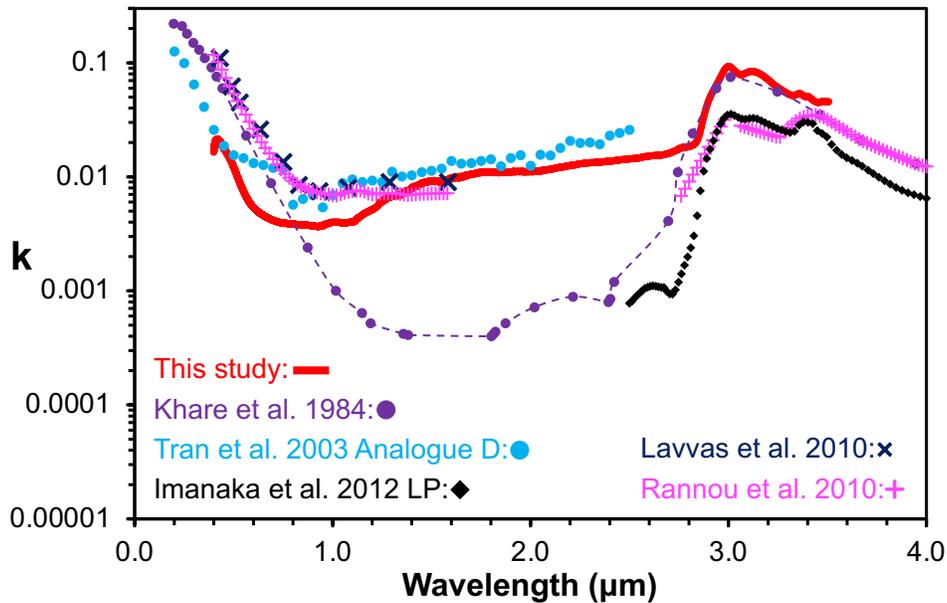

Figure 5. The extinction coefficients (k) retrieved from observational data (Lavvas et al. 2010; Rannou et al. 2010) and derived from laboratory-measured data (this study; Khare et al. 1984; Tran et al. 2003; Imanaka et al. 2012).

As shown in Figure 5, the optical constants from our current study and other studies can provide a better fit to the observations in certain wavelength range than Khare's data (Brassé et al. 2015). For example, the extinction coefficients (k) of sample D from Tran et al. (2003) are similar to the retrieved extinction coefficients (k) from the Cassini-Huygens observations in the 0.8-1.2 μm range, and our k values are closer to the retrieved ones from 1.2 to 1.6 μm (Lavvas et al. 2010; Rannou et al. 2010). Also, the 23-mbar sample from Imanaka et al. (2012) captures the broad absorption feature at 3.4 μm in agreement with observational data (Bellucci et al. 2009; Rannou et al. 2010). None of the lab measured optical constants can match the retrieved values from observations in the entire wavelength range. The discrepancy may reflect the compositional difference of Titan hazes at different altitudes, because various instruments on Cassini-Huygens spacecraft were sensitive to and probed different altitudes of Titan's atmosphere (Hörst 2017). Experiments with different setups and conditions simulate diverse chemical processes and result in tholins with distinct compositions, which may be representative of Titan hazes at different altitudes. Therefore, it is necessary to produce tholin samples under a variety of conditions ($CH_4/N_2$ ratios, pressures, and energy sources/densities) and determine their optical constants in a wide wavelength range. Doing so can help understand the formation and properties of Titan hazes. The measured optical constants will be useful not only for reanalyzing the observational data of Titan from past missions, but also for interpreting future Titan observations with the James Webb Space Telescope and the Dragonfly mission.

With developments in space exploration, proper optical constants of haze analogs for other planetary bodies are required. Due to the lack of relevant data, the optical data of Titan tholin from Khare et al. (1984) have been used for modeling Pluto (e.g., Zhang et al. 2017), Triton (e.g., Ohno et al. 2020), and even exoplanets (e.g., Arney et al. 2017; Adams et al. 2019; Gao & Zhang 2020). As we discussed earlier, the tholin sample in Khare et al. (1984) was prepared under specific conditions for simulating Titan's atmospheric chemistry, and the Khare's data do not adequately fit Titan observations. We should be cautious when applying the Khare's data to other bodies, especially to exoplanets. This study demonstrates an avenue for our future work to determine optical constants of haze analogs to other planetary bodies, such as Pluto, Triton, and various exoplanets (Hörst et al. 2018; He et al. 2018a, b; 2020a, b).

4. CONCLUSIONS

In this study, we prepared a Titan tholin film by exposing a $CH_4/N_2$ (5%/95%) gas mixture to cold plasma energy source at ~100 K with the PHAZER chamber, and determined the optical constants in the wavelength range from 0.4 to 3.5 μm using a vacuum FTIR. Due to the differences in experimental techniques, experimental conditions and measurement methods among different groups, it is difficult to strictly compare the optical constants of those tholins. In general, the optical constants of our tholin sample fall within the range of the reported values in the shorter wavelength (0.4 to 2.5 μm). In the range of 2.5 to 3.5 μm, our n values are the lowest while our k values are highest. The k values of our tholin sample are very close to the retrieved k values from the observational data (DISR and VIMS) in the 1.2-1.6 μm range. None of the tholin samples can meet the observational constraints of the Titan haze materials in the entire observed wavelength range, but tholins produced under different conditions may represent hazes formed at different altitudes in Titan's atmosphere. More detailed investigation is needed to understand the effect of the experimental conditions on the chemistry, and the connection between the chemical compositions and the optical properties of laboratory haze analogs. Such study will provide better constrains to Titan's haze materials and help interpret future Titan observations.

Acknowledgements

This work was supported by the NASA Astrophysics Research and Analysis Program NNX17AI87G and the NASA Exoplanets Research Program 80NSSC20K0271.